\documentclass{article}
\usepackage[utf8]{inputenc}

\usepackage{booktabs}
\usepackage{graphicx}
\usepackage{pifont}
\usepackage{authblk}

\title{Personality Traits and Ego-network Dynamics}

\author[1,2]{Simone Centellegher}
\author[3,4]{Eduardo L\'opez}
\author[5]{Jari Saram{\"a}ki}
\author[2]{Bruno Lepri}
\affil[1]{Department of Information Engineering and Computer Science, University of Trento, Trento, Italy}
\affil[2]{Mobile and Social Computing Lab, Fondazione Bruno Kessler (FBK), Trento, Italy}
\affil[3]{CABDyN Complexity Center, Saïd Business School, University of Oxford, Oxford, United Kingdom}
\affil[4]{Department of Computational and Data Sciences, College of Science, George Mason University, Fairfax, Virginia, US}
\affil[5]{Department of Computer Science, Aalto University School of Science, Espoo, Finland }
\date{}                     
\begin{document}

\maketitle

\section*{Abstract}

Strong and supportive social relationships are fundamental to our well-being. However, there are costs to their maintenance, resulting in a trade-off between quality and quantity, a typical strategy being to put a lot of effort on a few high-intensity relationships while maintaining larger numbers of less close relationships. It has also been shown that there are persistent individual differences in this pattern; some individuals allocate their efforts more uniformly across their networks, while others strongly focus on their closest relationships. Furthermore, some individuals maintain more stable networks than others. Here, we focus on how personality traits of individuals affect this picture, using mobile phone calls records and survey data from the Mobile Territorial Lab (MTL) study. In particular, we look at the relationship between \emph{personality traits} and the (i) \emph{persistence} of social signatures, namely the similarity of the social signature shape of an individual measured in different time intervals; (ii) the \emph{turnover} in egocentric networks, that is, differences in the set of alters present at two consecutive temporal intervals; and (iii) the \emph{rank dynamics} defined as the variation of alter rankings in egocentric networks in consecutive intervals. We observe that some traits have effects on the stability of the social signatures as well as network turnover and rank dynamics. As an example, individuals who score highly in the Openness to Experience trait tend to have higher levels of network turnover and larger alter rank variations. On broader terms, our study shows that personality traits clearly affect the ways in which individuals maintain their personal networks.

\section*{Introduction}
We interact with a wide network of people on a daily basis, and these social relationships play an important functional role in our lives. A large number of studies has shown that having strong and supportive relationships is essential for health and subjective well-being \cite{holtLunstad2010,lyubomirkyKingDiener2005}. As an example, the quantity and the quality of our social relationships reduce the risk of mortality \cite{berkmanSyme1979,house1988,kiecolt-glaserNewton2001}. Interestingly, this finding holds even when health behaviors, socioeconomic status, and other variables that might influence mortality are taken into account. 
Moreover, people experience more positive affect, one of the main components of subjective well-being, when they feel more connected to others \cite{ryff1989,reis2000}. Very happy people spend more time with family and friends and report more satisfying personal relationships with others, compared to people who are only moderately happy \cite{dienerSeligman2002}. At the same time, diversity in social interactions appears to lead to or correlate with desirable outcomes such as better health \cite{cohen1997,cohen2009}, positive affect, \cite{sandstrom2014} and higher level of creativity \cite{perry-smith2006}.\\

However, there are also costs to maintain close and diverse relationships and it has been shown that the interactions and relationships a subject (\emph{ego}) has with family members and friends (\emph{alters}) may be subject to general constraints associated with time available for interactions \cite{winship1978,dunbar1992,hansen2001,southerton2003} and human cognitive abilities to interact with a large number of alters \cite{miller1956,stiller2007,powell2012}. Recently, the increasing availability of data on human communication has opened enormous opportunities for uncovering the mechanisms governing time allocation in social networks \cite{onnela2007,miritello2013limited,miritello2013time,alshamsi2016network} in a way that circumvents biases typical to retrospective self-reports \cite{eagle2009,lazer2009}. In line with previous sociological findings \cite{winship1978,dunbar1992,hansen2001,southerton2003}, these studies show that, in general, individuals mostly interact with a small subset of their personal network, and that the effects of time constraints grow with the network size: individuals with large networks tend to dedicate, on average, less time to each relationship than people who have small social networks \cite{goncalves2011,miritello2013limited,miritello2013time}. 

In recent work, Saram{\"a}ki \emph{et al.} \cite{saramaki2014persistence} used auto-recorded mobile phone data to investigate the way egos divide their communication efforts (calls) among alters and how persistent the observed patterns are over time. They show that individuals display a distinctive and robust \emph{social signature} that captures how phone call interactions are distributed across different alters. Interestingly, they find evidence that for a given ego these signatures tend to persist over time, despite a considerable turnover in the identity of alters.\\ 

In the present paper, we bring individual dispositions such as personality traits into the picture. In particular, our aim is to investigate whether personality traits of individuals are associated with their communication patterns in the form of social signatures. Scientific psychology defines the notion of personality traits as stable dispositions towards action, belief and attitude formation. Hence, personality traits are relatively stable over time, different across individuals (e.g. some people are outgoing whereas others are shy), and play an important role in influencing people behaviour \cite{matthews2009sustained,costa1992four}. However, several studies have shown that personality traits do not exist in a vacuum and traits are meaningful only if they are considered together with situations in the generation of behavior \cite{funder2006}. Specifically, such situations encompass all the environmental input that we experience, including the physical environment and all the living beings we interact with. A large proportion of what makes situations relevant for people is the interaction with other people \cite{diener1984}.\\

For example, Staiano \emph{et al.} \cite{staiano2012friends} considered the role of a number of structural ego-network metrics (e.g. centrality measures, triads, efficiency, transitivity) in the prediction of personality traits, using self-assessments as a ground truth. An interesting finding is the tendency of extroverts to keep their close partners together, also by promoting their introduction to each other. Using social data from Facebook and more precisely from the ego-networks containing the list of ego's friends, Friggeri \emph{et al.} \cite{friggeri2012} found a negative correlation between Extraversion and the partition ratio. The partition ratio quantifies the extent to which the communities of an ego-network are disjointed from one other. Hence, this result implies that individuals with high scores in Extraversion tend to be in groups that are linked to each other, while individuals with low scores in Extraversion tend to
be in more distinct and separate social groups. This observation is compatible with the results obtained by
Staiano \emph{et al.} \cite{staiano2012friends} showing the extroverts' tendency of introducing friends belonging to different communities.
In another study using data from Facebook, Quercia \emph{et al.} \cite{quercia2012} studied the relationship between Facebook popularity (number of contacts) and personality traits on a large number of individuals. They found that popular users (those with many social contacts) tend to have high scores in Extraversion and low scores in Neuroticism. In particular, they found that the Extraversion score is a good predictor for the number of Facebook contacts.

In this work, we focus on understanding whether and how personality traits affect the (i) \emph{persistence} of social signatures, namely the similarity of the social signature shape of an individual measured in different time intervals; (ii) the \emph{turnover} in egocentric networks, that is, differences in the set of alters present at two consecutive temporal intervals; and (iii) the \emph{rank dynamics} defined as the variation of alter rankings in egocentric networks in consecutive intervals.

Specifically, we combine detailed mobile phone call records with personality traits scores collected from survey data. The mobile phone calls records were collected during the Mobile Territorial Lab (MTL) study \cite{centellegher2016mobile}, and tracked the daily communication patterns of more than 100 parents over a period of two years. In the current work, we use the communication networks of 93 individuals over a period of 15 months.\\

On broader terms, our study shows that personality traits clearly affect the ways in which individuals maintain their personal networks. Specifically our results show that extroverts tend to show slightly lower temporal persistence of their social signatures, as compared to introverts.
Moreover, people with high scores in the Openness to Experience personality trait exhibit a higher network turnover with respect to their counterpart and interestingly agreeable individuals have a lower turnover inside their network of alters than disagreeable ones. 
In addition we found that changes in the intensity of relationships result in increased or decreased communication with alters, which is reflected in the alter rank dynamics inside ego-networks. We found a larger variation in the alters' ranks of egos who show higher scores in the Openness to Experience personality trait, and the opposite for egos who show lower scores in the same trait. This is also true for the Agreeableness personality trait. 

\section*{Methods}

\subsection*{Procedure}
In the current study, we leverage the sensing technologies available in smartphones and track the daily communication networks of 93 individuals in Trento, Italy, for a period of 15 months. The study was conducted within the Mobile Territorial Lab (MTL), a joint living lab created by Telecom Italia, Fondazione Bruno Kessler, MIT Media Lab and Telef\'onica. Following Italian regulations, all participants were asked to sign an informed consent form and the study was conducted in accordance to it. The general study and the form were also approved by a joint Ethical Committee of University of Trento and Province of Trento.\\

The MTL living lab consists of a group of more than 100 volunteers who carry an instrumented smartphone in exchange for a monthly credit bonus of voice, SMS and data access. The sensing system installed on the smartphones is based on the FunF framework \cite{aharony2011} and keeps track of communication events through call and SMS logs, the user’s location thanks to the GPS sensor and the location of the cell towers the phone is connected to, the applications' usage and other kinds of useful information. One of the most important features of such a living lab is its ecological validity, given that the participants' behaviors and attitudes are sensed in the real world, as people live their everyday lives.\\

All volunteers were recruited within the target group of young families with children, using the snowball sampling approach, where existing study participants recruit future participants from
among their acquaintances \cite{goodman1961}. Upon joining the living lab, each participant filled out an initial questionnaire for collecting their demographics and information on individual traits and other dispositions.\\

\subsection*{Materials}
Self-assessment questionnaires have been used to measure the personality of each individual in terms of the Big Five model \cite{costa1992four}. This model comprises five personality traits: (i) Extraversion (sociable, assertive, playful vs. aloof, reserved, shy), (ii) Agreeableness (friendly, cooperative vs. antagonistic, faultfinding), (iii) Conscientiousness (self-disciplined, organized vs. inefficient, careless), (iv) Neuroticism (insecure, anxious vs. calm, unemotional), and (v) Openness to Experience (intellectual, insightful vs. shallow, unimaginative).

\begin{figure}[!h]
    \includegraphics[width=\textwidth]{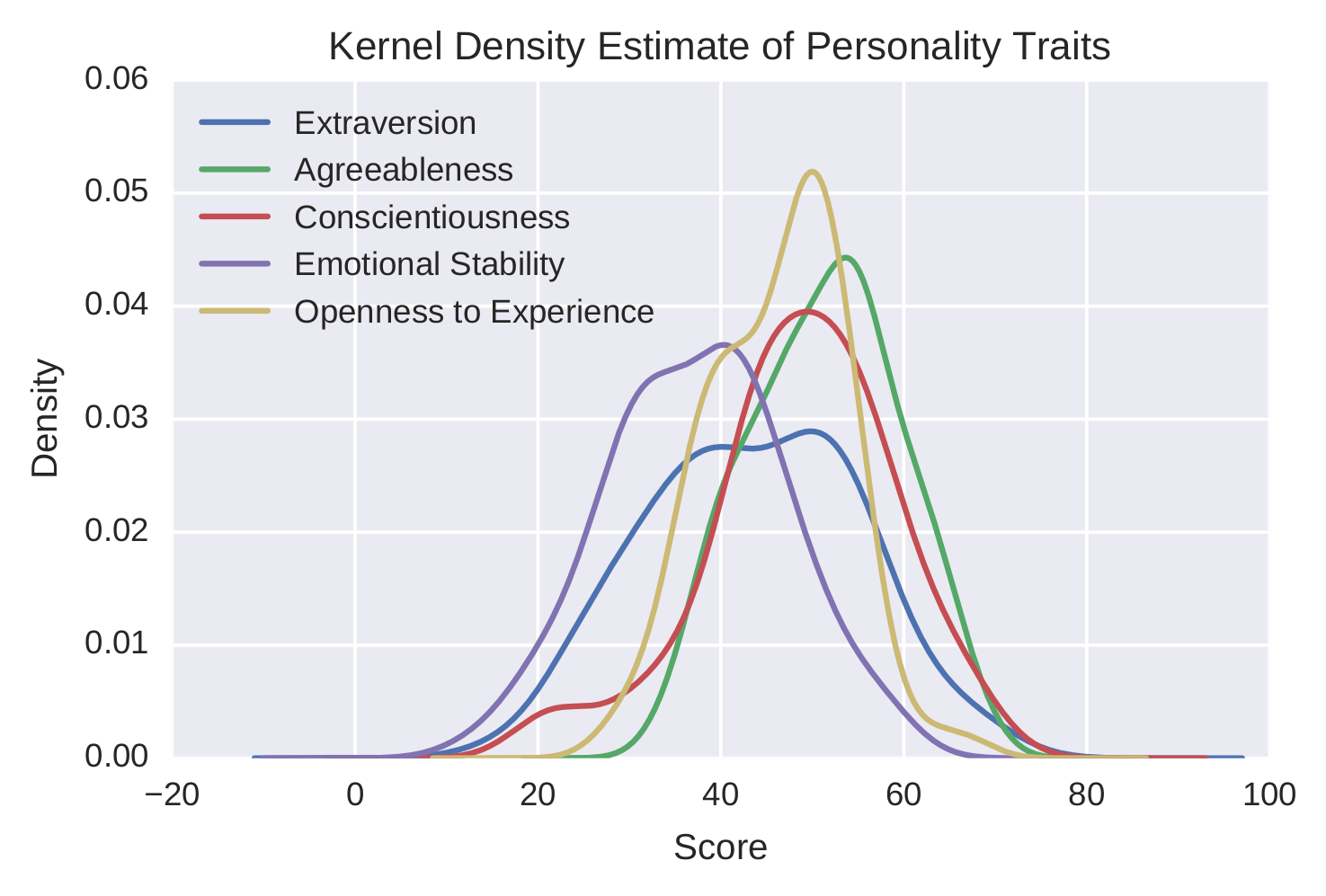}
    \caption{{\bf Kernel Density Estimate of the distribution of the Big Five Personality Traits.} The estimated probability density functions are computed using a non-parametric Gaussian kernel density estimator that employs Scott's rule of thumb for bandwidth selection.}
    \label{fig1}
\end{figure}

The Italian version of the Big Five Marker Scale (BFMS) \cite{perugini2002big} was used to assess the personality traits at the beginning of the experiment. This validated scale is an adjective list composed by 50 items, with personality scores between 15 and 70 (see Fig~\ref{fig1}).
For a detailed description of the Mobile Territorial Lab initiative refer to \cite{centellegher2016mobile}.\\

\subsection*{Data Preprocessing}
All analyses presented in the following sections are based on 15 months of data collected between October, 2013 and December, 2014. 
Social signatures are generated by following the method of Saram{\"a}ki \emph{et al.} \cite{saramaki2014persistence}, by counting the number of calls to each alter, ranking the alters by this count, and then computing the fraction of calls associated with each rank. 
In order to study the communication patterns of each individual, we use only the outgoing phone calls because they represent the effort made by an individual to maintain a particular social relationship. We divide the 15 month observation period in three intervals $I_{1}$, $I_{2}$ and $I_{3}$ of 5 months each.
We chose a 5 month interval because the entire period of 15 months was the period that allowed us to have the higher number of participants for a longer period of time. Moreover, choosing too short intervals, we could face the problem of mostly measuring fluctuations. Too long intervals would not work either, since social signatures would contain too many alters who have already left the network.
We retain all the participants that made at least 150 calls and contacted at least 20 people in each of the three intervals.
The result of this process leaves us with a set of 93 out of 142 participants, 56 females and 37 males. The participants' ages range from 28 to 48 years, with an average of 39 years.\\


First, following the assumption that individuals in the extreme of the scale for a given trait would exhibit largest differences in communication patterns, we identify for each of the Big Five personality traits
people falling in the 25th percentile (low personality scores) and the 75th percentile (high personality scores). Thus, for example, for the Extraversion trait we find the most extroverted individuals and the most introverted individuals. These groups of top and bottom scoring individuals will be used throughout the study for comparisons. The sizes of the groups are presented in Table~\ref{tab:sub_size}.

\begin{table}[!ht]
\centering
\begin{tabular}{@{}lcc@{}}
\toprule
\textbf{Personality Trait} & \textbf{25\% Sample (Low)} & \textbf{75\% Sample (High)} \\ \midrule
Extraversion & 23 & 23\\
Agreeableness & 22 & 23 \\
Openness to Experience & 22 & 16 \\
Conscientiousness & 20 & 23 \\
Emotional Stability & 19 & 21\\
\bottomrule
\end{tabular}
\caption{Personalities subgroups sizes of people falling in the 25th percentile (low personality scores), and people falling in the 75th percentile (high personality scores)}
\label{tab:sub_size}
\end{table}

\subsection*{Ego-network Dynamics}

\subsubsection*{Persistence}
In order to evaluate the shape similarity of two different social signatures, we used the Jensen-Shannon divergence (JSD):
\begin{equation}
JSD(P_1,P_2)= H\left ( \frac{1}{2}P_1 + \frac{1}{2}P_2 \right )-\frac{1}{2}[H(P_1)+H(P_2))]
\end{equation}
where $P_i=\{p_i(r)\}$ is a social signature and $p_i(r)$ represents the fraction of calls made by an ego to the alter of rank $r$ in signature $i$.  $H$ represents the Shannon entropy defined as
\begin{equation}
H(P)=-\sum_{r=1}^{k}p(r)\ log\ p(r)
\end{equation}
where $p(r)$ is defined as above and $k$ represents the total number of alters called by a particular ego. The lower bound of the JSD is zero and intuitively the lower the value of the JSD the more similar two signatures are.

\begin{figure}[!ht]
    \includegraphics[width=\textwidth]{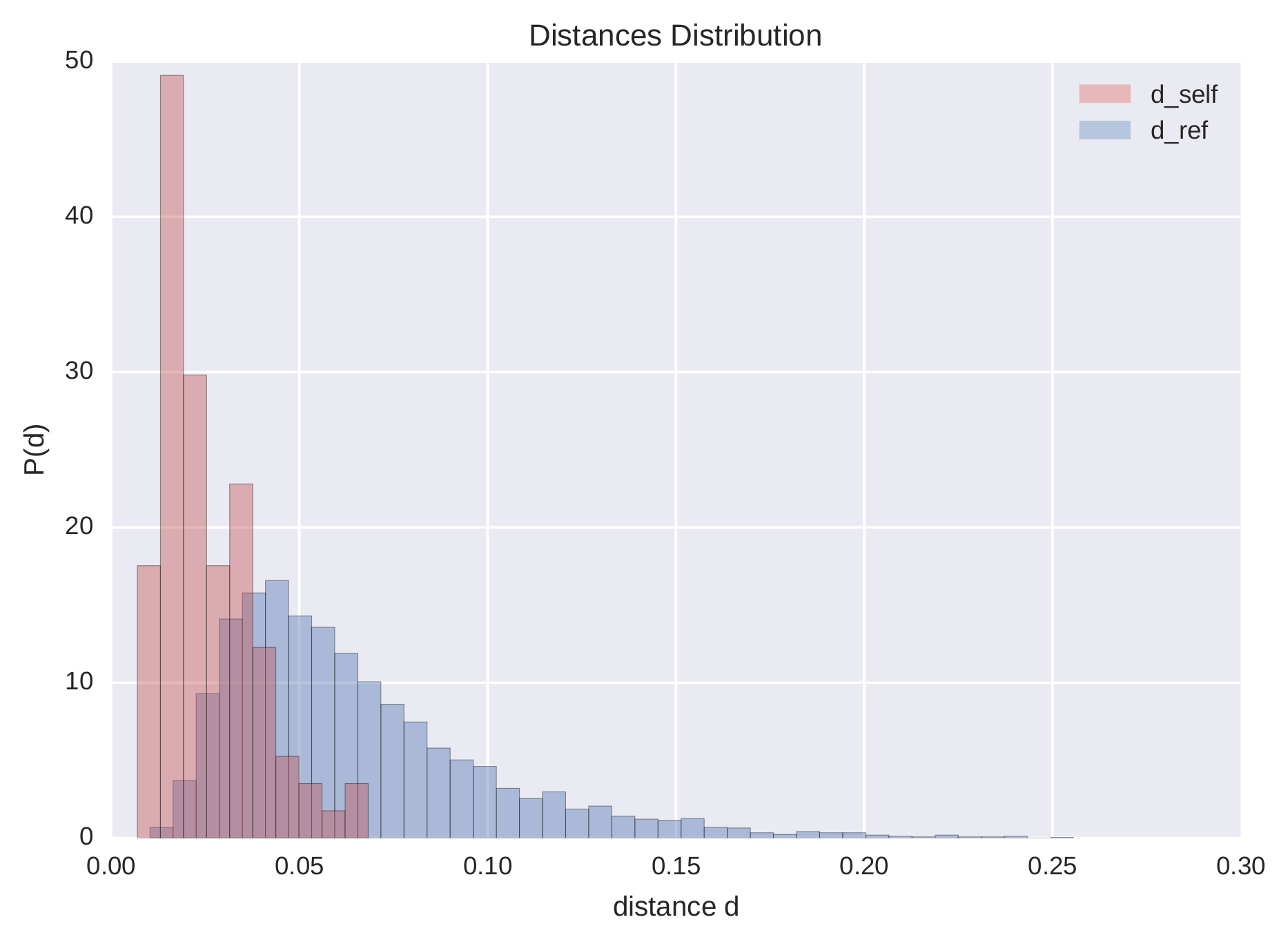}
    \caption{{\bf Self and reference distance distributions.} Distribution of self ($d_{self}$) and reference ($d_{ref}$) distances of the social signatures of the entire population in consecutive intervals, showing that the ego's signatures are typically similar with respect to the signatures of the other egos}
    \label{fig2}
\end{figure}

Following \cite{saramaki2014persistence} and using the JSD defined above, we computed the self distance $d_{self}$ for each ego, which quantifies the similarity of the ego's signatures in two consecutive intervals $(I_t, I_{t+1})$. We also computed reference distances $d_{ref}$ which quantify, for each interval, the similarity between the signature of a particular ego $i$ and the signatures of all other egos $j$. Fig~\ref{fig2} shows the distribution of the self and reference distances of the entire population under observation. These distributions are in line with the results in \cite{saramaki2014persistence} and indicate that individuals' signatures remain similar in shape in consecutive intervals.\\

\subsubsection*{Turnover}
\begin{figure}[!h]
    \includegraphics[width=\textwidth]{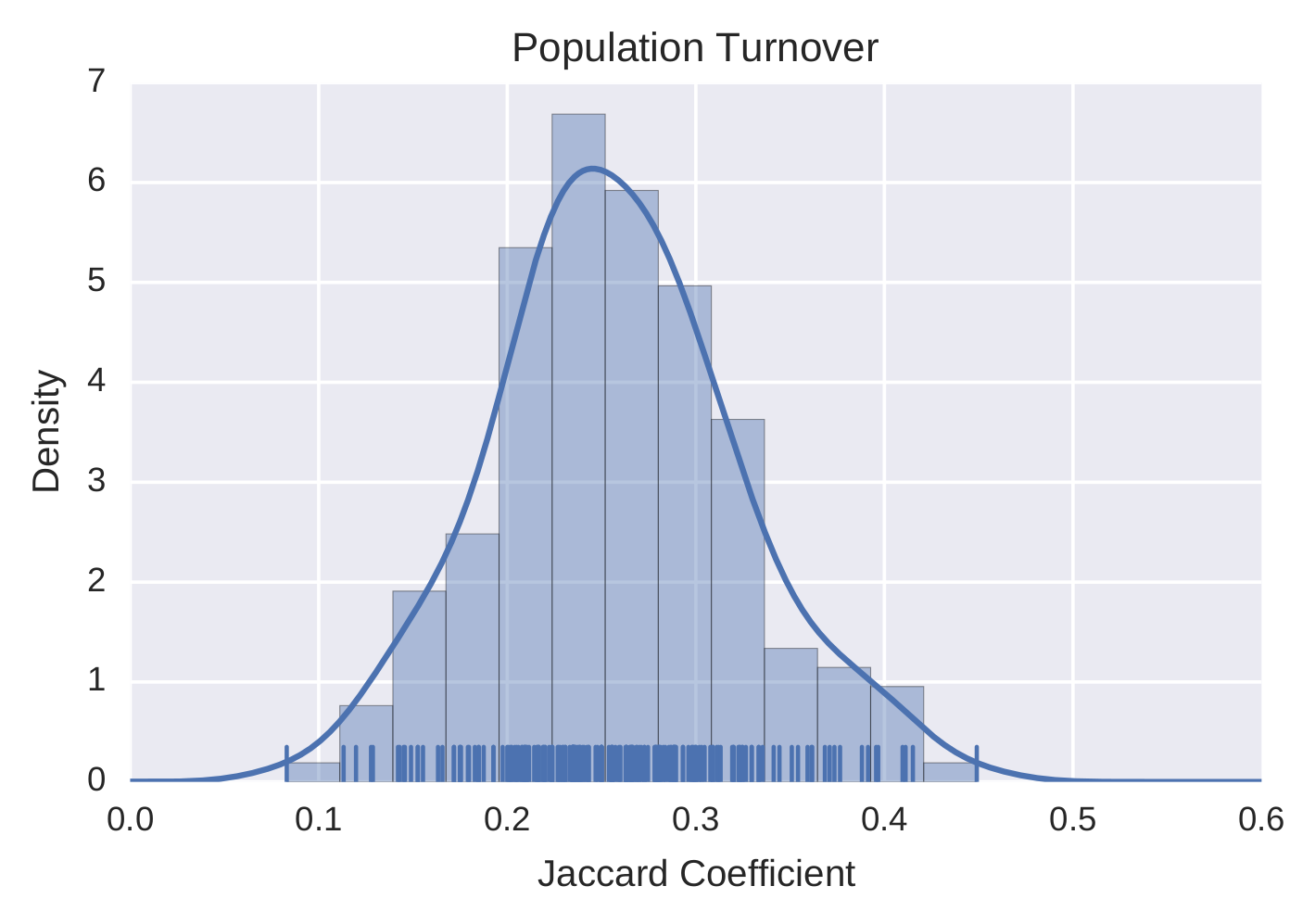}
    \caption{{\bf Population turnover distribution.} Turnover distribution inside the ego networks of the entire population for both ($I_1$,$I_2$) and ($I_2$,$I_3$). The average of the Jaccard similarity coefficient is $\langle J \rangle=0.257$, showing that on average there is an high turnover between ego networks in two consecutive intervals. The lower the Jaccard index, the higher the turnover. The estimated probability density function of the sample is computed using a non-parametric Gaussian kernel density estimator that employs Scott's rule of thumb for bandwidth selection.}
    \label{fig3}
\end{figure}

The turnover inside each ego network, namely the differences between the sets of alters present in two consecutive intervals, is measured with the Jaccard similarity coefficient as
\begin{equation}
J(I_i,I_j) = \frac{|A(I_i)\cap A(I_{j})|}{|A(I_i)\cup A(I_{j})|}
\end{equation}
where $A(I_i)$ and $A(I_j)$ represent the set of alters called by a particular ego in time intervals $I_i$ and $I_j$, respectively.
Fig~\ref{fig3} shows the distribution of turnover for the ego networks of the 93 people under observation ($\langle J \rangle=0.257$).

\section*{Results}
In this section we present the results of our analysis on personality traits and ego-network dynamics. Typically, when looking at different aspects of the social signatures of the 25th and 75th percentile subgroups for a given trait, we find that their distributions do not follow a normal distribution. Therefore, in order to assess if there are significant differences between the distributions of the two opposite subgroups we apply two statistical tests: (1) the non-parametric Kruskal-Wallis test to verify whether the population medians of the two subgroups are equal, and (2) the non-parametric Kolmogorov-Smirnov test to verify whether the cumulative distribution functions of the two subsets are identical.

\subsection*{Personality traits and ego-network size}

We first evaluate whether personality traits have some effect on the ego-network size. For each subgroup, we find that the distribution of network sizes is right skewed (positive skewed). We use the network size of the subgroups in each of the three intervals. In Table~\ref{tab:net_size_stat} we report the median, the first quartile (Q1) and the third quartile (Q3) for each subgroup and the results with a significance level of $p < 0.05$; for these personality traits, network sizes are significantly different for the 25th and 75th percentiles, that is the trait does have an effect on network size. 

\begin{table}[!ht]
\centering
\begin{tabular}{@{}llrrrrr@{}}
\toprule
 &  &\textbf{Median} & \textbf{Q1} & \textbf{Q3} & \textbf{KW} & \textbf{KS} \\ \midrule
\textbf{Openness to Experience} & \textbf{High} & 75.5 & 53.8 & 89.0 &  &\\
 & \textbf{Low} & 86.0 & 66.0 & 114.0 & 4.74* & \\
\midrule
\textbf{Extraversion}& \textbf{High} & 79.0 & 60.0 & 113.0 &  &\\
& \textbf{Low} & 71.0 & 57.0 & 90.0 &  &\\
\midrule
\textbf{Agreeableness}& \textbf{High} & 80.0 & 61.0 & 95.0 &  & \\
& \textbf{Low} & 66.0 & 54.0 & 84.0 & 6.51* & 0.29**\\
\midrule
\textbf{Conscientiousness}& \textbf{High} & 78.0 & 57.0 & 92.0 & & \\
& \textbf{Low} & 67.0 & 48.5 & 84.0 &  &\\
\midrule
\textbf{Emotional Stability}& \textbf{High} & 84.0 & 60.5 & 112.5 &  & \\
& \textbf{Low} & 79.0 & 57.0 & 99.0 & & \\ \bottomrule
\end{tabular}
\caption{Statistics for egocentric network sizes of different trait subgroups. The median, the first quartile (Q1) and the third quartile (Q3) for each subgroup are reported. We performed the Kruskal-Wallis test (KW) and the Kolmogorov-Smirnov test (KS) in order to assess eventual differences between the distributions of the reference distances of opposite subgroups (e.g. extroverts and introverts). Only the Kruskal-Wallis (KW) and Kolmogorov-Smirnov (KS) tests with $p < 0.05$ are reported. Note: * $p<0.05$, ** $p<0.01$, *** $p<0.001$.}
\label{tab:net_size_stat}
\end{table}

The median values of the network size distribution of subgroups of people with high and low scores in the Agreeableness personality trait, show statistically significant differences, with median network sizes of 80.0 and 66.0, respectively.
The subgroups of people with high and low scores in the Openness to Experience trait have a median network size of 75.5 and 86.0, respectively, but they show a significant statistical difference only with the Kruskal-Wallis test.
Non significant differences are found between the subgroups of the other three personality traits (Extraversion, Conscientiousness and Emotional Stability).

\subsection*{Personality traits and the persistence of social signatures}
Here, we try to understand the relationship between the persistence of a social signature and the Big Five personality traits. We investigate whether the self-distances of subgroups of opposite personality traits (e.g. extroverts and introverts) exhibit differences, which would indicate that the signatures are more persistent for one group than for the other.
We thus try to understand whether a particular personality disposition influences the stability of an individual signature over time.

We find a significant difference only in the distributions of the self-distances of the subgroups of extroverts and introverts, namely people with high and low scores in the Extraversion personality trait: the signatures of extroverts are less persistent than the signatures of introverts (see Table~\ref{tab:persistence_diff_self}). However, it is worth noticing that only the Kruskall-Wallis test shows a significant statistical difference while the Kolmogorov-Smirnov test does not.

\begin{table}[!ht]
\centering
\begin{tabular}{@{}lllllll@{}}
\toprule
 & & \multicolumn{1}{c}{\textbf{Median}} & \multicolumn{1}{c}{\textbf{Q1}} & \multicolumn{1}{c}{\textbf{Q3}} & \multicolumn{1}{c}{\textbf{KW}} & \multicolumn{1}{c}{\textbf{KS}} \\ \midrule
\textbf{Openness to Experience} & \textbf{High} & 0.021 & 0.017 & 0.041 &  &  \\
& \textbf{Low} & 0.018 & 0.015 & 0.026 &  &  \\
\midrule
\textbf{Extraversion}& \textbf{High} & 0.022 & 0.019 & 0.034 & 5.27* &  \\
& \textbf{Low} & 0.018 & 0.016 & 0.022 &  &  \\
\midrule
\textbf{Agreeableness}& \textbf{High} & 0.022 & 0.018 & 0.034 &  &  \\
& \textbf{Low} & 0.025 & 0.015 & 0.035 &  & \\
\midrule
\textbf{Conscientiousness}& \textbf{High} & 0.020 & 0.014 & 0.033 &  &  \\
& \textbf{Low} & 0.022 & 0.017 & 0.036 &  &  \\
\midrule
\textbf{Emotional Stability}& \textbf{High} & 0.022 & 0.017 & 0.033 &  &  \\
& \textbf{Low} & 0.019 & 0.017 & 0.024 &  & \\ \bottomrule
\end{tabular}
\caption{Self-distances of social signatures within subgroups. The median, the first quartile (Q1) and the third quartile (Q3) for each subgroup are reported. We performed the Kruskal-Wallis test (KW) and the Kolmogorov-Smirnov test (KS) in order to assess eventual differences between the distributions of the self distances of opposite subgroups (e.g. extroverts and introverts). Note: * $p<0.05$, ** $p<0.01$, *** $p<0.001$.}
\label{tab:persistence_diff_self}
\end{table}

\subsection*{Turnover}
\begin{figure}[!h]
    \includegraphics[width=\textwidth]{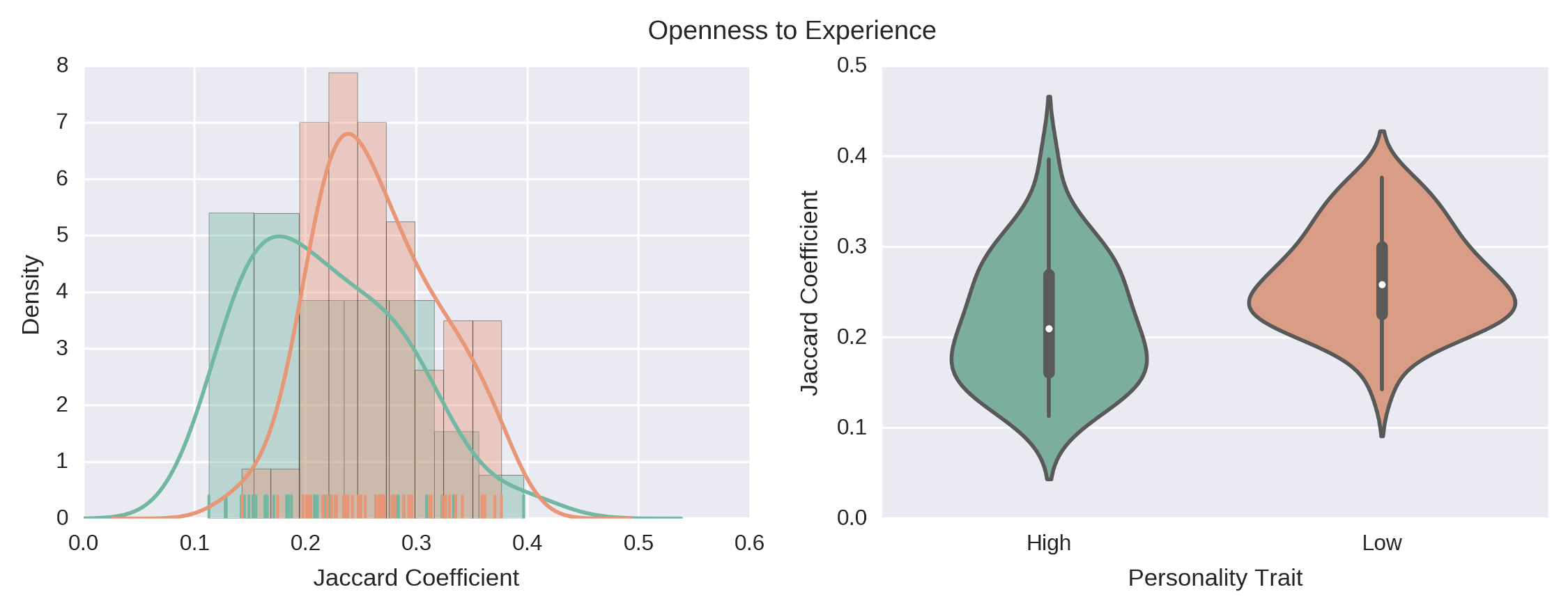}
    \caption{{\bf Openness to Experience and network turnover.} Individuals who are more open to experience show higher turnover, with a median value of 0.21, as compared to the lowest-scoring 25\% who show a median turnover value of 0.26. Left: the estimated probability density functions are computed using a non-parametric Gaussian kernel density estimator that employs  Scott's rule of thumb for bandwidth selection. Right: violin plots of the same distributions.}
    \label{fig4}
\end{figure}

We also investigated the association between personality traits and the turnover in ego-networks in two consecutive intervals. Again, we use the Kruskal-Wallis and the Kolmogorov-Smirnov tests (see Table~\ref{tab:turnover_diff}). As shown in Fig~\ref{fig4}, our results show that network turnover tends to be characterized by the Openness to Experience trait, where people that are willing to try new experiences exhibit a higher network turnover (median = 0.210) with respect to people who are more closed to experience (median = 0.259).

\begin{table}[!ht]
\centering
\resizebox{\textwidth}{!}{
\begin{tabular}{@{}lllllll@{}}
\toprule
 & & \multicolumn{1}{c}{\textbf{Median}} & \multicolumn{1}{c}{\textbf{Q1}} & \multicolumn{1}{c}{\textbf{Q3}} & \multicolumn{1}{c}{\textbf{KW}} & \multicolumn{1}{c}{\textbf{KS}} \\ \midrule
\textbf{Openness to Experience} & \textbf{High} & 0.210 & 0.161 & 0.270 & 9.31** & 0.39** \\
& \textbf{Low} & 0.259 & 0.226 & 0.300 &  &  \\
\midrule
\textbf{Extraversion}& \textbf{High} & 0.253 & 0.209 & 0.312 &  &  \\
& \textbf{Low} & 0.265 & 0.230 & 0.295 &  &  \\
\midrule
\textbf{Agreeableness}& \textbf{High} & 0.279 & 0.237 & 0.323 & 12.76*** & 0.384** \\
& \textbf{Low} & 0.235 & 0.204 & 0.264 &  &  \\
\midrule
\textbf{Conscientiousness}& \textbf{High} & 0.266 & 0.228 & 0.317 &  &  \\
& \textbf{Low} & 0.237 & 0.210 & 0.284 &  &  \\
\midrule
\textbf{Emotional Stability}& \textbf{High} & 0.267 & 0.225 & 0.316 &  &  \\
& \textbf{Low} & 0.270 & 0.218 & 0.298 &  &  \\ \bottomrule
\end{tabular}
}
\caption{Network turnover as measured by the Jaccard coefficient for the different subgroups. The median, the first quartile (Q1) and the third quartile (Q3) for each subgroup are reported. We performed the Kruskal-Wallis test (KW) and the Kolmogorov-Smirnov test (KS) in order to assess differences between the distributions of the turnover inside the ego networks of opposite subgroups (e.g. extroverts and introverts). The subgroups in the top-25\% for the Openness to Experience and the Agreeableness traits show higher turnover with respect to their opposite personality trait subgroups. Note: * $p<0.05$, ** $p<0.01$, *** $p<0.001$.}
\label{tab:turnover_diff}
\end{table}

\begin{figure}[!h]
    \includegraphics[width=\textwidth]{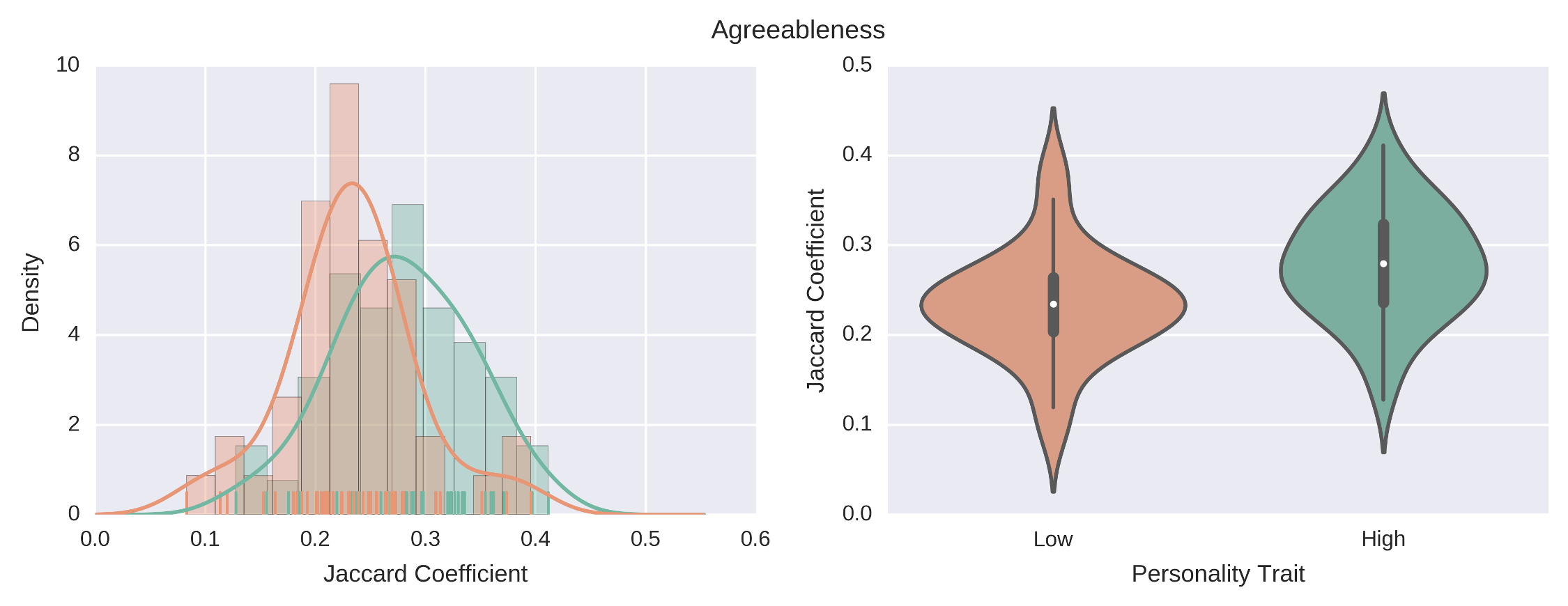}
    \caption{{\bf Agreeableness and network turnover.} People with low scores in the Agreeableness trait, thus more disagreeable people, show a higher turnover, with a median value of 0.23, with respect to more agreeable people who show a median turnover value of 0.28. Left: The estimated probability density functions are computed using a non-parametric Gaussian kernel density estimator that employs the Scott's rule of thumb for bandwidth selection. Right: Violin plots of the same distributions.}
    \label{fig5}
\end{figure}

Network turnover seems to be characterized by the Agreeableness personality trait as well. Fig~\ref{fig5} shows that generally more likable people have a lower network turnover as compared to disagreeable individuals. This could be considered counter-intuitive if one expects that an agreeable person would be more social and therefore s/he would communicate with a more diverse set of people. On the other hand, a reasonable-sounding explanation is that people having difficulties getting along with others are less likely to have a stable set of alters, and they probably struggle in having long-term relationships with a lot of people, resulting in a higher network turnover. 

Finally, we do not find any significant differences for Extraversion, Conscientiousness and Emotional Stability.

\subsection*{Rank Dynamics}
In the previous section, we have seen that the Openness to Experience and the Agreeableness traits associate with network turnover. Here, we take a detailed look at what happens inside the network of a focal ego by focusing at the alters rank dynamics and subsequently we analyze the effect of personality traits on such dynamics.
To this end, for two consecutive temporal intervals for each ego, we build a transition matrix $A$ as follows: if there is a transition of an alter from rank $i$ in interval $I_t$ to rank $j$ in interval $I_{t+1}$, then $A_{ij} = 1$.
We limit the maximum rank to 20, because this guarantees that the population of 93 individuals has an alter at each rank in each 5-month interval.

We also introduce a row labelled $i$ (21st row) to represent the probability for alters inside an ego network to enter ranks 1-20 from beyond the maximum considered rank of 20 in the next time interval. The row labelled $in$ (22nd row) is then introduced to represent the probability for a new alter to join the ego network in the next time interval. The $o$ (21st) and $on$ (22nd) columns represent the probability of moving beyond the 20th rank or completely dropping out of the network, respectively.

In this way, the transition matrix of each ego keeps track of rank dynamics of alters and also the dynamics of alters exiting or entering the network. 

We then used the transition matrices of egos to represent the alter rank variations of entire subgroups. To this end, we simply sum the matrices of all egos in the subgroup and normalize them by the number of egos in that particular subgroup, in order to have  probabilities on both rows and columns. The resulting matrix now contains the alters rank dynamics represented as probabilities of moving up and down rank positions. We call this resulting matrix $B$.

\begin{figure}[!h]
    \includegraphics[width=\textwidth]{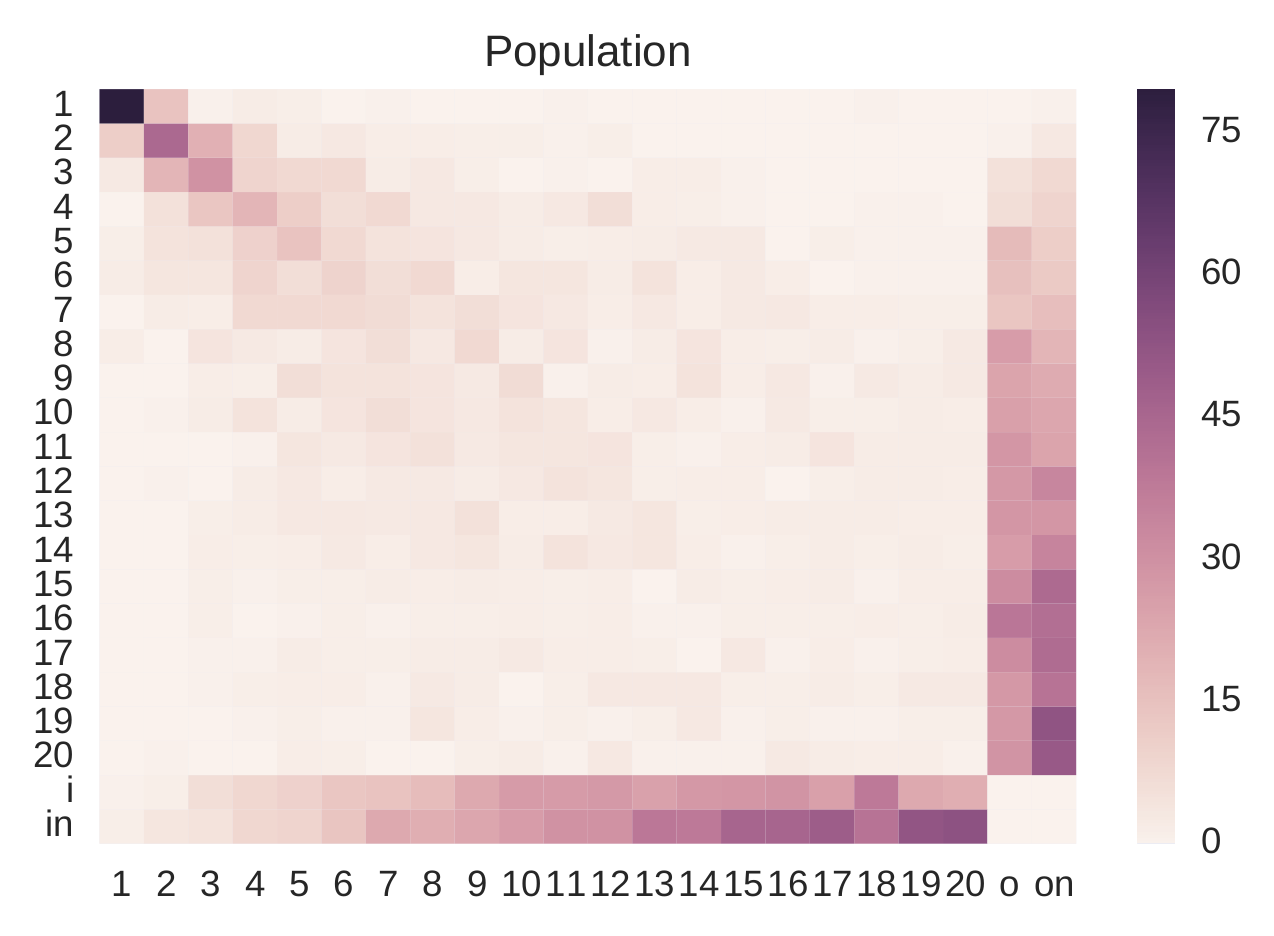}
    \caption{{\bf The normalized transition matrix for the entire population.} The row labelled $i$ represents the probability for alters beyond the maximum rank of 20 to move up to a more central position in the next time interval. The row labelled $in$ represents the probability for a new alter to join the network in the next time interval. The $o$ and $on$ columns represent the probability of moving out beyond the 20th position or completely dropping out of the network, respectively. Looking at the diagonal of the transition matrix, it is possible to notice that the top positions are more stable with respect to low-ranked positions.}
    \label{fig6}
\end{figure}

Fig~\ref{fig6} shows the normalized transition matrix $B$ of the entire population in both ($I_1$,$I_2$) and ($I_2$,$I_3$).

For the top ranks, the probability mass is clearly concentrated on the diagonal, meaning that the top ranks are more stable. This is expected, since people in the top positions of the network are the people that a particular ego contacts more frequently, such as for example family members, and these relationships are expected to be more close and stable. Also notice that approximately beyond the 10th rank, alters have a higher probability to drop out of the network with respect to higher-ranked alters (columns $o$ and $on$), while it is easier to enter the network to lower-rank positions (columns $i$ and $in$).\\

Next, we investigated whether personality traits affect the stability of the ego-network. We quantify the network stability \cite{wasserman1994social,sales2007extracting} in the following way:
\begin{equation}\label{eq:dist}
C = \frac{1}{N}\sum_{i=1}^{N}\sum_{j=1}^{N} B_{ij}|i-j|.
\end{equation}
This measure calculates for each element $B_{ij}$ of the transition matrix $B$ the distance of the element from the diagonal and then averages over all values. If $C = 0$, all ego-networks in the consecutive intervals $I_t$ and $I_{t+1}$ are exactly the same, as no alters change their ranks. Intuitively, the more stable a network is, the more ``heat'' will be concentrated on and near the diagonal. In contrast, the more unstable the network, the more spread-out the ``heat'' of the transition matrix will be. Note that the definition of Eq~\ref{eq:dist} does not include the special rows/columns $i, in, o, on$.

\begin{figure}[!h]
    \includegraphics[width=\textwidth]{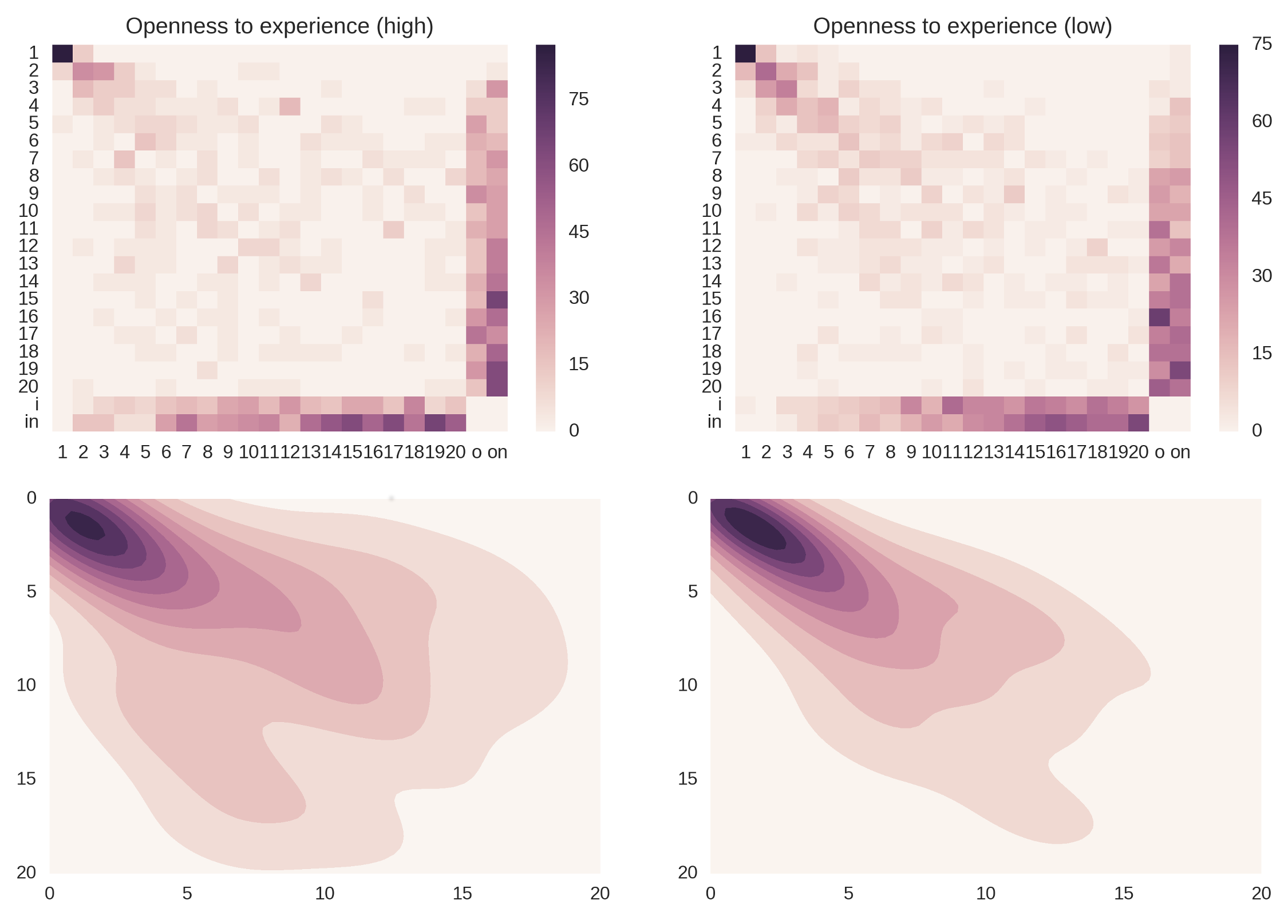}
    \caption{{\bf Rank dynamics for the Openness to Experience trait.} Top row: the transition matrices for the subgroups of individuals with high and low scores in the Openness to Experience personality trait. It is possible to observe that the subgroup of people that display higher scores ($C = 0.452$) shows a higher spread with respect to the opposite subgroup ($C = 0.383$), where the ``heat'' is more concentrated around the diagonal. Also the columns/rows that represent the alters that fall out/in the 20th position or the network show a higher stability in the subgroup of people with low scores (their values increase more slowly when moving towards higher ranks). Bottom row: The 2-dimensional kernel density estimation plots emphasize the fact the rank variations inside the group of people with high scores in the Openness to Experience trait are larger with respect to the opposite subgroup. The estimated probability density functions are computed using a non-parametric Gaussian kernel density estimator that employs Scott's rule of thumb for bandwidth selection.}
    \label{fig7}
\end{figure}

Fig~\ref{fig7} shows the transition matrices of the subgroups of individuals with high Openness to Experience scores and individuals with low Openness to Experience scores. 

As it is possible to see, the subgroup of people with high scores in the Openness to Experience personality trait seems less stable than the subgroup with the opposite personality disposition. This is also clearly observable in the corresponding 2-dimensional kernel density estimation plots. Applying Eq~\ref{eq:dist}, the subgroup of people that have higher scores, namely people more open to experience, has a network stability values of $C = 0.452$ and the subgroup of people with low scores has a value of $C = 0.383$. It seems that people that show a higher disposition to curiosity and willingness to experiment new things tend to be less stable regarding the set of alters that they communicate with. 
In order to check the validity of these results, we also calculated the distance from the diagonal for the neutral group of individuals that display neither high nor low scores in the Openness to Experience trait; these should represent the ``middle ground'' between the extremes, and therefore their stability value should fall between the values of the highest- and lowest-scoring groups. This is indeed the case, as the neutral group exhibited a distance value of $C = 0.443$.\\

\begin{figure}[!h]
    \includegraphics[width=\textwidth]{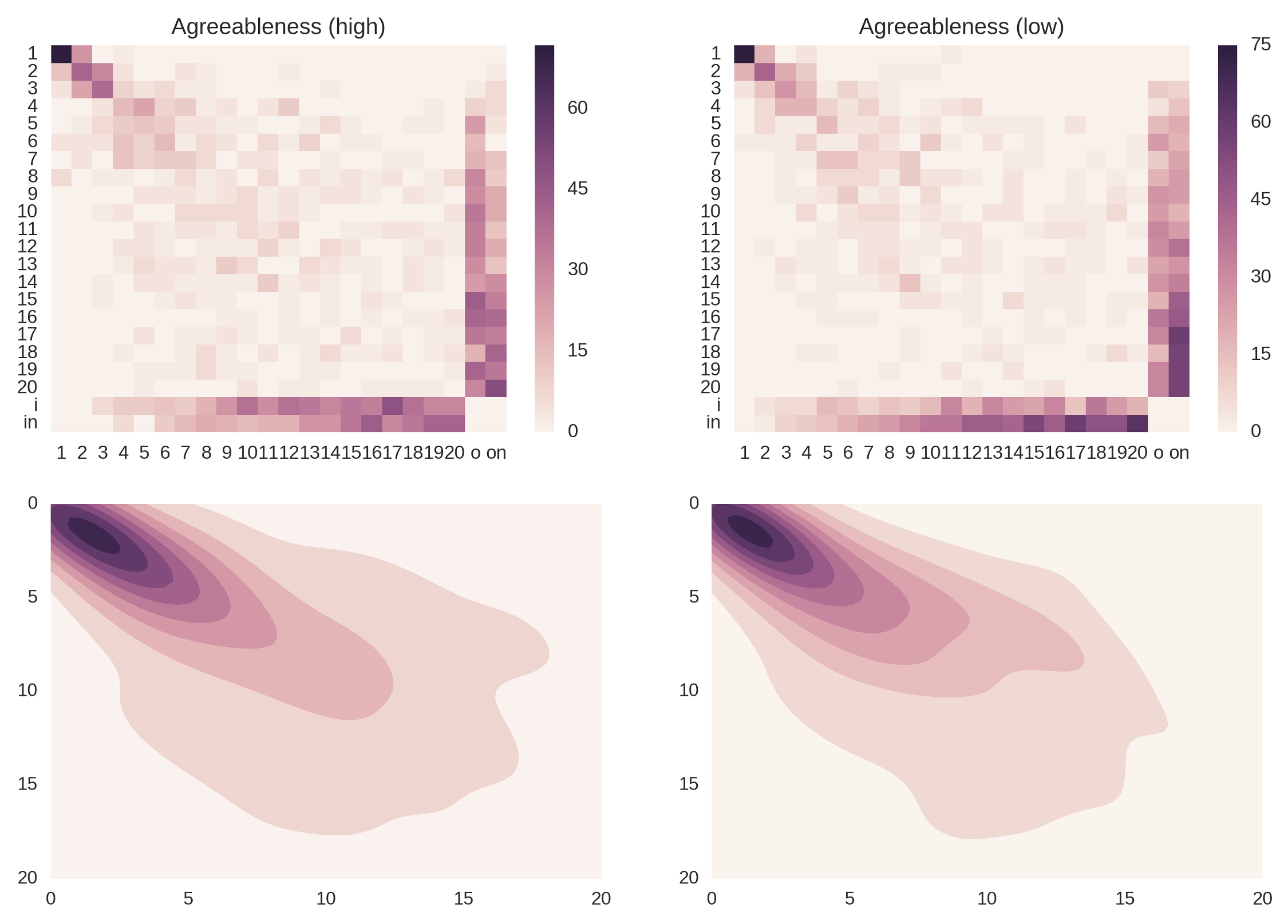}
    \caption{{\bf Rank dynamics for the Agreeableness trait.} Top row: transition matrices for the subgroups of individuals with high and low scores in the Agreeableness personality trait. The subgroup of agreeable people, namely the group of people with high scores ($C = 0.461$), shows an higher spread with respect to the subgroup of people with low scores ($C =  0.373$), where the ``heat'' is more concentrated on the diagonal. Bottom row: the 2-dimensional kernel density estimation plots emphasize the fact the rank variations inside the subgroup of people with high scores in the Agreeableness trait are larger with respect to the opposite subgroup. The estimated probability density functions are computed using a non-parametric Gaussian kernel density estimator that employs  Scott's rule of thumb for bandwidth selection.}
    \label{fig8}
\end{figure}

We have similar results with the Agreeableness personality trait. More agreeable people tend to have a higher spread, namely larger rank dynamics with respect their counterpart, as shown in Fig~\ref{fig8}. The distance $C$ for the subgroups of individuals with high scores, low scores and the middle group for the Agreeableness personality trait are 0.461, 0.373 and 0.441, respectively.\\

We do not detect significant differences for the other Big Five personality traits, including, surprisingly, the Extraversion trait.

\section*{Discussion}
In this paper, we have investigated if and how personality traits are related to the ways people allocate their communication across the members of their social network. Specifically, we focused on differences in \emph{social signatures} and their persistence between opposite subgroups of a particular trait, as well as the stability of their ego-networks in terms of turnover and rank dynamics of alters.

Our results show that some personality traits play a role in characterizing the \emph{persistence} of the social signatures, as well as the \emph{turnover} and the \emph{rank dynamics} of ego-networks (see Table~\ref{tab:results}). 

\begin{table}[!ht]
\centering
\resizebox{\textwidth}{!}{
\begin{tabular}{@{}lccc@{}}
\toprule
 & \multicolumn{1}{c}{\textbf{Persistence ($d_{self}$)}} &  \multicolumn{1}{c}{\textbf{Turnover}} & \multicolumn{1}{c}{\textbf{Rank Dynamics}} \\ \midrule
\textbf{Openness to Experience} &  & \ding{51} & \ding{51}\\
\textbf{Extraversion} & \ding{51}$^{+}$ &  &  \\
\textbf{Agreeableness} &  & \ding{51} & \ding{51} \\
\textbf{Conscientiousness} &  &   & \\
\textbf{Emotional Stability} &  &   & \\
\bottomrule
\end{tabular}
}
\caption{Results summary. The table shows aspects that seem to be affected by the Big Five personality traits. Note that the results highlighted with the $^{+}$ symbol are significant only with the Kruskall-Wallis test}
\label{tab:results}
\end{table}

We find that extroverts tend to show slightly lower temporal persistence of their social signatures, as compared to introverts.

People willing to try new experiences (as indicated by high scores in the Openness to Experience personality trait) exhibit a higher network turnover with respect to their counterpart. Interestingly, agreeable individuals have a lower turnover inside their ego-networks than disagreeable ones. In social psychology, Agreeableness and Extraversion are the traits having the most direct implications for social interactions \cite{tov2014}. However, the two traits, although positively correlated, reflect distinct implications. Extraverts have been described as assertive, talkative, and motivated to engage in social contact \cite{wiltRevelle2009}. In contrast, agreeable people are characterized as likable and concerned with maintaining positive relationships with others \cite{graziano2009,tobin2000}, as also confirmed in our analyses by their tendency in investing in longer and more stable communication relationships.

As said, Openness to Experience and Agreeableness have an impact on the turnover inside the ego network of an individual, and partially explain why new alters are added to the network and why old alters are replaced. In addition to this turnover, changes in the intensity of relationships may result in increased or decreased communication with alters, which is reflected in the ranks dynamics inside ego-networks. We found a larger variation in the alter ranks of egos who display higher scores in the Openness to Experience personality trait, and the opposite in the subgroup of people with low scores.
The Agreeableness trait also affects turnover: more agreeable people have a lower network turnover and thus longer relationships as compared to their counterpart. However, those more agreeable exhibit ego networks with larger alter rank variations. A possible explanation is given by the fact that agreeable people are more likable and easy going and thus they do not display preferences in adapting their behaviour to alters with respect to their counterpart.
Therefore, it seems that these two personality traits play a relevant role in the rank dynamics of the ego networks.

Turning to the limitations of the present study, we list the
following ones: the relatively small size of the sample; the
fact that it comes from a group of young families with children living in the same geographical area (Trento, in the northern of Italy); the non-availability of data from different communication channels such as Whatsapp, FaceTime, etc. 

Despite these limitations, our results overall provide one possible explanation for the uniqueness and stability of the individuals' social signatures. As pointed out by Saram\"aki \emph{et al.}, \cite{saramaki2014persistence} social signatures' characteristics reflect the fact that ego networks are typically layered into a series of hierarchically inclusive subsets of relationships of different quality. One of the constraints shaping the social signatures seems to be 
the one arising from differences in personality traits, with some individuals preferring to have a few, intense, and stable relationships and others preferring more diverse, but less intense ones \cite{swickert2002}. 

However, additional constraints, such as available time \cite{roberts2011,miritello2013limited} and cognitive limits \cite{bernard1973,powell2012}, may influence the unique pattern represented by an individual’s social signature. It is also possible that there are factors that influence social signatures in combination (\emph{e.g.}~joint effects of multiple personality traits together with other drivers). Determining these will require more detailed studies and access to different kinds of data.

\end{document}